\documentclass{ws-mpla}

\begin{document}

\markboth{S. Rahvar, M. Sadegh Movahed and M Saadat} {The Effect
of Uncertainty Principle on the Thermodynamics of Early Universe}

%
\catchline{}{}{}{}{}
%

\title{The Effect of Uncertainty Principle on the Thermodynamics of Early
Universe }

\author{Sohrab Rahvar}

\address{Department of Physics, Sharif University of Technology,\\
 Tehran, P.O. Box 11365-9161, Iran\\
 Institute for Studies in Theoretical Physics and Mathematics,\\
 Tehran, P.O.Box 19395-5531, Iran\\
rahvar@sharif.edu}

\author{Mohammad Sadegh Movahed}

\address{Department of Physics, Sharif University of Technology,\\
 Tehran, P.O. Box 11365-9161, Iran\\
 Institute for Studies in Theoretical Physics and Mathematics,\\
 Tehran, P.O.Box 19395-5531, Iran\\
m.s.movahed@mehr.sharif.edu}

\author{Mehdi Saadat}

\address{Department of Physics, Sharif University of Technology,\\
 Tehran, P.O. Box 11365-9161, Iran\\
 msaadat@sharif.edu}
\maketitle

\pub{Received (Day Month Year)}{Revised (Day Month Year)}

\begin{abstract}
We discuss the concept of measurement in cosmology from the
relativistic and quantum mechanical points of view. The
uncertainty principle within the particle horizon, excludes the
momentum of particles to be less than $\pi\hbar H/c$. This effect
modifies the standard thermodynamics of early universe for the
ultra-relativistic particles such that the equation of state as
well as dependence of energy density and pressure to the
temperature. We show that this modification to the thermodynamics
of early universe is important for energies $E>10^{17} GeV$.
During the inflation, the particle horizon inflates to a huge size
and makes the uncertainty in the momentum to be negligible.
\keywords{Cosmology; Foundations of quantum mechanics; measurement
theory; Quantum statistical mechanics}
\end{abstract}
\begin{figure}
\begin{center}
\includegraphics[angle=0,scale=0.4]{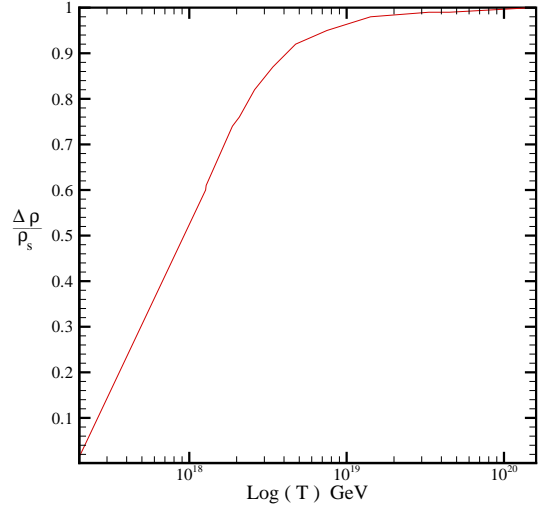}
\caption{ \label{dc} Relative difference between the standard
energy density $\rho_s$ and that of considering the uncertainty
correction $\rho$, $(\frac{\rho_s - \rho}{\rho_s})$ as a function
of $\log(T)$ in GeV.}
\end{center}
\end{figure}

\ccode{PACS Nos.: 98.80.-k, 03.65.Ta, 05.30.-d}
 \vspace{1cm}
The concept of measurement is the basis of two main theories of
physics, relativity and quantum mechanics. At the present universe
where the cosmological scales are too large and energy density of
cosmological fluid is too low, the measurement can be treated
classically. However, for the early universe with higher energies
and smaller scales, using quantum mechanical concept in the
measurement is essential.\\
As we know from the basis of quantum mechanics, if we confine a
particle inside a box, an uncertainty due to this confinement
appears in the momentum space. A well known example is the
scattering of the particles from the slit, where the confinement
of the position causes uncertainty in the momentum, resulting the
formation of the scattering pattern. Analogue to a particle inside
a box, in the case of the early universe we have a causal box
(i.e. particle horizon) which any observer in the universe is
confined to do his/her measurements within this scale \cite{wald}.
In another word considering a wave function to a particle, the
probability of finding a particle by an observed
outside its horizon is zero, $\Psi^2(\lambda>horizon)= 0$.\\
From the theory of relativity, measurement of a stick length can
be done by sending simultaneous signals to the observer from the
two endpoints, where for the scales larger than the causal size,
those signals need more than the age of the universe to be
received. Looking back to the history of the universe, the
particle horizon after the Planck era grows as $cH^{-1}$ but
inflates to a huge size by the beginning of inflationary epoch
\cite{bran}. In the pre-inflationary epoch the maximum uncertainty
in the location of a particle $cH^{-1}$ results in a minimum
uncertainty in the momentum as:
\begin{equation}
p_{min} = \pi \hbar H/c,
\end{equation}
where $H$ is the Hubble parameter.\\
This uncertainty can also be described by an experiment similar to
the Heisenberg Gedanken microscope. Using photons with small
wavelengths to probe the position of a particle makes larger
uncertainty in the momentum. To minimize this effect, we have to
use photons as large as possible wavelengths. However photons with
the wavelengths larger than the particle horizon are not
detectable. The horizon size wavelength photons are the weakest
accessible probe, producing a minimum uncertainty of $p =\pi\hbar
H/c$. As we go back further to the Planck epoch the size of
particle horizon becomes smaller and consequently the
uncertainly in momentum raises.\\
For the ultra-relativistic particles the minimum uncertainty of
energy in three spatial dimensions is $E_{min} =\sqrt{3}\pi\hbar
H$. To have a feeling from the amount of uncertainty in the energy
we compare it with the thermal energy through ${E_{min}}/{kT}$.
This ratio grows as we go further to the early universe. In the
Planck epoch $kT \simeq M_{pl}$ and $H^2 \simeq {M_{pl}}^2$, so
these two energies tend to the same order of magnitude. \\
Excluding the momenta smaller than $P_{min}=\pi \hbar H/c$ from
the phase space, modifies the standard thermodynamics of early
universe. We start with the partition function for the fermions
and bosons to calculate thermodynamical parameters;
\begin{equation}
\ln Z = \pm\sum_n\ln(1\pm e^{-\beta E_n}),
\end{equation}
where $\beta = {1}/{kT}$ and upper and lower signs stand for the
fermions and bosons, respectively. For the ultra-relativistic
particles with the continuum energy spectrum, the partition
function can be written as:
\begin{eqnarray}
\ln Z  &=& \pm g \int_{E_{min}}^{\infty}\frac{4\pi
n^2}{8}\ln(1\pm e^{-\beta E_n})dn,\nonumber \\
&=&\mp \frac{gV}{6\pi^2 \hbar^3 c^3}E_{min}^3\ln(1\pm e^{-\beta E_{min}})\nonumber \\
&+&\frac{\beta gV}{6\pi^2\hbar^3
c^3}\int_{E_{min}}^{\infty}\frac{E^3dE}{e^{\beta E}\pm 1},
\end{eqnarray} where $E_n = n\pi\hbar H$, ${4\pi n^2}/{8}$ in the
first integrand is the density of states, $g$ is the degree of
freedom for a quantum state and $V$ is the volume of the causal
box. We use standard thermodynamics, $P=\frac{kT}{V}\ln Z$ to
calculate the pressure as follows:
\begin{equation}
P =\frac{\rho c^2}{3}\mp \frac{g}{6\pi^2 \hbar^3 c^3
\beta}E_{min}^3\ln(1\pm e^{-\beta E_{min}}), \label{press}
\end{equation}
where the first term at the right hand side of equation of state
comes from the energy density as:
\begin{equation}
\rho
=\frac{g}{2\pi^2\hbar^3c^5}\int_{E_{min}}^{\infty}\frac{E^3dE}
{e^{\beta E}\pm 1}, \label{density}
\end{equation}
and the second term is the correction term to the pressure, say it
$P_c$. Fig. (\ref{per}) shows the deviation of Eq. (\ref{press})
from the standard definition of pressure in terms of
$E_{min}/{KT}$. Close to the Planck epoch $E_{min}/{KT}
\rightarrow 1$ and the deviation from the standard definition of
pressure, $P_c/P$ is $\sim 5.4 \%$ for the fermions and $\sim 7\%$
for the bosons.\\
The dependence of the mass density of fluid of fermions and bosons
to the temperature also can be obtained according to the Eq.
(\ref{density}) as:
\begin{eqnarray}
\label{f} \rho_{\it  f}&=&\frac{7g_f\pi^2K^4T^4}{240c^5\hbar^3}
-\frac{g_f}{240c^5\hbar^3\pi^2}[30E_{min}^4 -7\pi^4K^4T^4\nonumber\\
&-& 120E_{min}^3KT\log(1+e^{\frac{E_{min}}{KT}}) +360E_{min}^2K^2T^2\nonumber\\
&\times&\sum_{n=1}^{\infty}\frac{e^{\frac{nE_{min}}{KT}}}{n^2}
-720E_{min}K^3T^3\sum_{n=1}^{\infty}\frac{e^{\frac{nE_{min}}{KT}}}{n^3}\nonumber\\
&+&720K^4T^4\sum_{n=1}^{\infty}\frac{e^{\frac{nE_{min}}{KT}}}{n^4}]. \\
\rho_{\it b}&=&\frac{g_b\pi^2K^4T^4}{30c^5\hbar^3}
-\frac{g_b}{120c^5\hbar^3\pi^2}[-15E_{min}^4 -4\pi^4K^4T^4\nonumber\\
&+& 60E_{min}^3KT\log(1-e^{\frac{E_{min}}{KT}})+180E_{min}^2K^2T^2\nonumber\\
&\times&\sum_{n=1}^{\infty}\frac{e^{\frac{nE_{min}}{KT}}}{n^2}
-360E_{min}K^3T^3\sum_{n=1}^{\infty}\frac{e^{\frac{nE_{min}}{KT}}}{n^3}\nonumber\\
&+&360K^4T^4\sum_{n=1}^{\infty}\frac{e^{\frac{nE_{min}}{KT}}}{n^4}].
\label{b}
\end{eqnarray}
If we let $E_{min} = 0$, we will get the familiar relation between
the mass density and temperature for the ultra-relativistic
fermions and bosons \cite{pady}. The density deviation from the
standard thermodynamics $\frac{\rho_s - \rho}{\rho_s}$ is shown in
Fig. (\ref{dc}). For temperatures near the Planck epoch, the
density is lower than that of standard one. For the entropy
density calculation $s =
-\frac{1}{V}\times{\partial{F}}/{\partial{T}}$ we use the free
energy $F = -kT\ln Z$ which results in:
\begin{equation}
s= \frac{4\rho c^2}{3T} \mp \frac{g}{6\pi^2 \hbar^3
c^3\beta}E_{min}^3\ln(1\pm e^{-\beta E_{min}}).
\end{equation}
The second term is the correction to the standard thermodynamics
and grows as we approach to the early universe.\\
The equation of state can be obtained by substituting Eqs.
(\ref{press}) and (\ref{density}) in the continuity equation
$\dot{\rho} + 3\frac{\dot a}{a}(\rho + P/c^2) =0 $, as follows:
\begin{figure}
\begin{center}
\includegraphics[angle=0,scale=0.4]{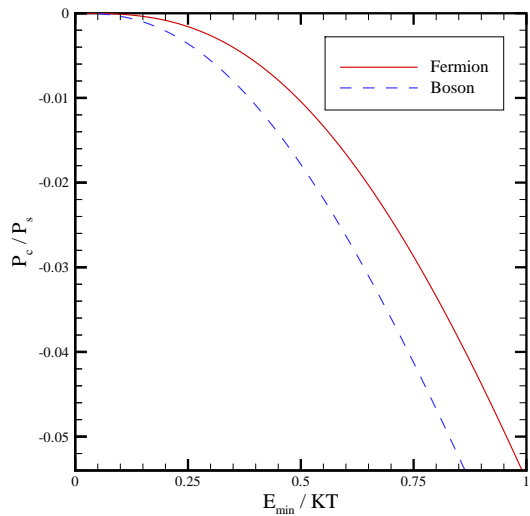}
\caption{\label{per} The ratio of the correction term of the
pressure to the of standard definition of the pressure as a
function of $E_{min}/{kT}$. For the low energies, this ratio tends
to zero.}
\end{center}
\end{figure}

\begin{figure}
\begin{center}
\includegraphics[angle=0,scale=0.4]{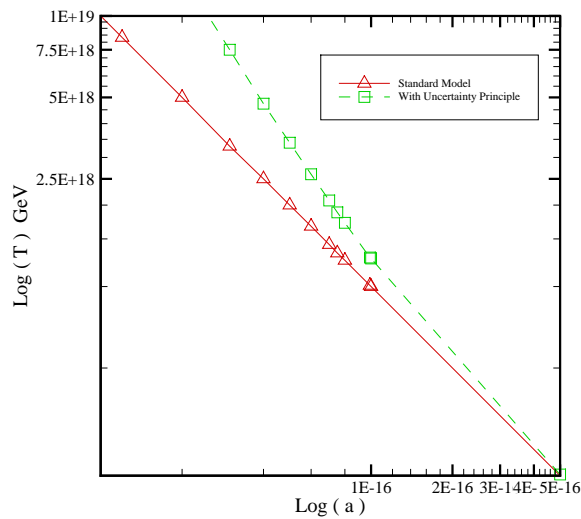}
\caption{ \label{t_ver_a} The temperature as a function of the
scale factor, in logarithmic scales. The solid line indicates this
dependence in the normal thermodynamics of universe $(T\propto
1/a)$ and the dashed line is for the case that we use the
uncertainty principle. Here the scale factor $a=1$ is chosen for
the electroweak energy.}
\end{center}
\end{figure}
\begin{eqnarray}
\frac{d\rho}{da} &=& -\frac{3}{a}(\rho + P/c^2), \nonumber \\
&=& -\frac{3}{a}(\frac{4}{3}\rho + P_{c}/c^2).
\end{eqnarray}
We use the iteration method and apply the standard thermodynamics
to solve the continuity equation up to the first order correction
for the fermions and bosons as:
\begin{eqnarray}
\label{rhoa} \frac{d \varrho_f}{da}
&=& -\frac{4}{a}\left[\varrho_f - 46.8 g^{3/4}a^{-6}{\varrho_f}^{1/4}\right], \\
\frac{d \varrho_b}{da} &=& -\frac{4}{a}\left[\varrho_b - 389.6
ga^{-8}\right], \label{rhob}
\end{eqnarray}
where $\varrho$ is normalized to the Planck energy density. The
analytical solutions of Eqs. (\ref{rhoa}) and (\ref{rhob}) are:
\begin{eqnarray}
\label{sa}
\varrho_f &=& \left[ a^{-3} + 46.8 g^{3/4}(a^{-3} - a^{-6})\right]^{3/4}, \\
\varrho_b &=& a^{-4} + 389.6 g(a^{-4} - a^{-8}). \label{sb}
\end{eqnarray}
Here we have deviation from the standard dependence of density to
the scale factor as $\rho\propto a^{-4}$. To obtain the dependence
of the scale factor to the temperature we use the solutions of the
Eqs. (\ref{sa}) and (\ref{sb}) at the left hand side of Eqs.
(\ref{f}) and (\ref{b}) and ignore the second order corrections.
The deviation of temperature from $1/a$ for the fermions and
bosons is shown in Fig. (\ref{t_ver_a}), where it behaves as
$1/a^{n}$ with the index of $n\simeq 1.5$.\\
In summery we have shown that a natural constraint to the lower
limit of particles energy appears due to their confinement at the
particle horizon. The effect is the modification of the
thermodynamics of the very early universe for energies $E>10^{17}
GeV$. In consequence, the equation of state and dependence of
temperature, pressure, energy density to the scale factor are all
modified. Inflating the size of the particle horizon during the
inflation, makes the uncertainty of the momentum to be negligible.
This effect may shed light on the thermodynamics of early universe
considering the cosmic fluid generated from the energy momentum of
quantum fields \cite{davis}.
\section*{Acknowledgments}
The authors thank Mohammad Nouri-Zonoz for reading the manuscript
and giving useful comments.


\begin{thebibliography}{0}    

\bibitem{wald} R. M. Wald, {\it General Relativity} (The University
of Chicago Press, 1984).

\bibitem{bran} R. H. Brandenberger in {\it Large Scale Structure Formation},edited by
R. Mansouri and R. Brandenberger (Kluwer Academic Publishers,
2000).

\bibitem{pady} T. Padmanabhan, {\it Structure Formation in the
Universe} (Cambridge university press, 1993).

\bibitem{davis} N. D. Birrell, P. C. W. Davies, P. V. Landshoff,{\it Quantum Fields
in Curved Space} (Cambridge Monographs on Mathematical Physics,
1982)


\end{thebibliography}
\end{document}